\documentclass[aps,prb,twocolumn,groupedaddress,showpacs,fleqn,floatfix]{revtex4}

\usepackage{graphicx}
\usepackage{epsfig}
\usepackage{dcolumn}
\usepackage{bm}
\usepackage{amsmath}
\usepackage{verbatim}
\usepackage[usenames]{color}

\newcommand{\figa} {
\begin{figure}
\centering
\includegraphics[width=.45\textwidth]{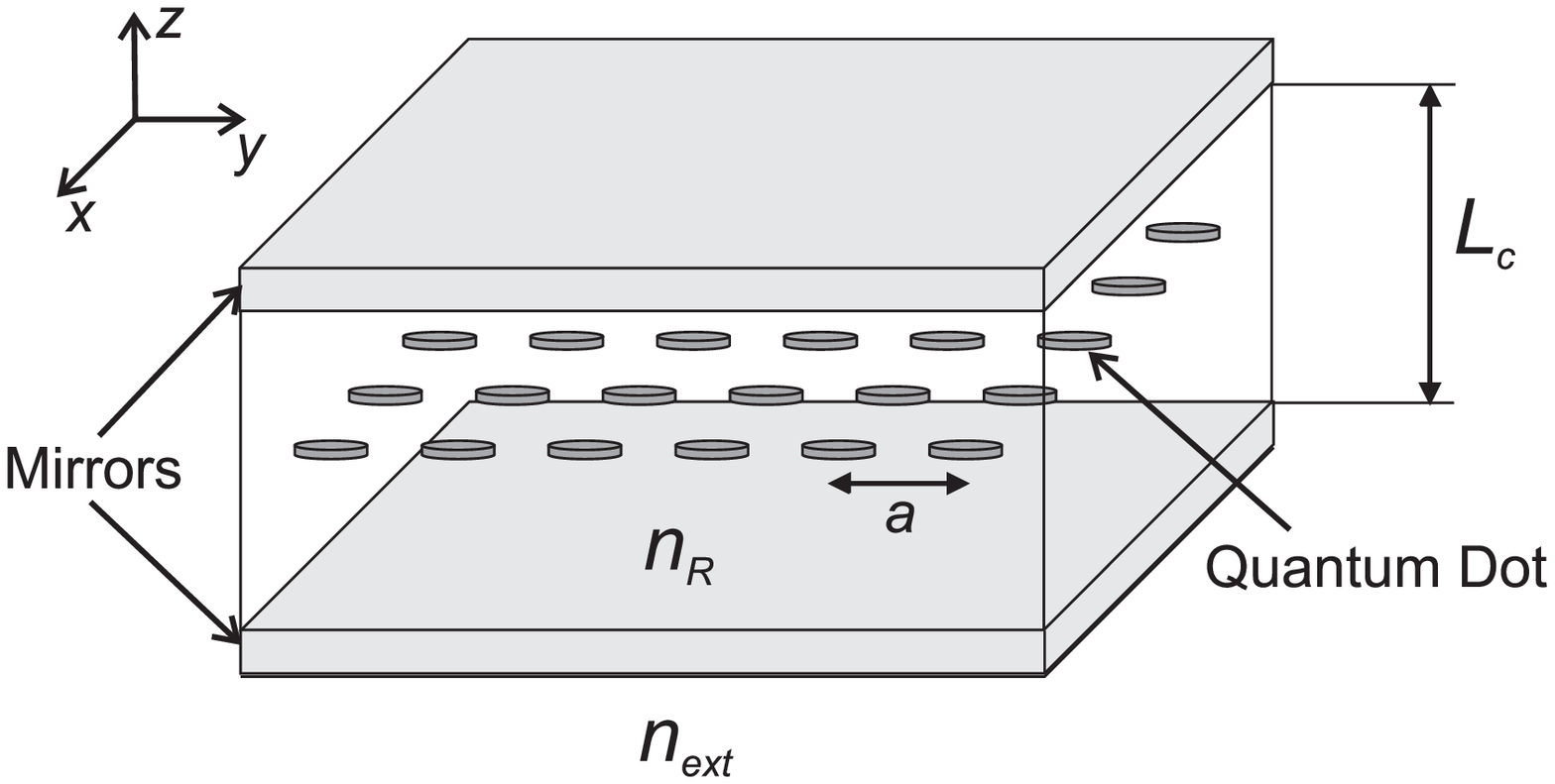} 
\caption{Scheme of the system studied. An ideal quantum dot lattice is
embedded in a planar microcavity of length $L_c$ and refractive index $n_R$.}
\label{figa}
\end{figure}
}

\newcommand{\figbb} {
\begin{figure}
\centering 
\includegraphics[width=0.4\textwidth]{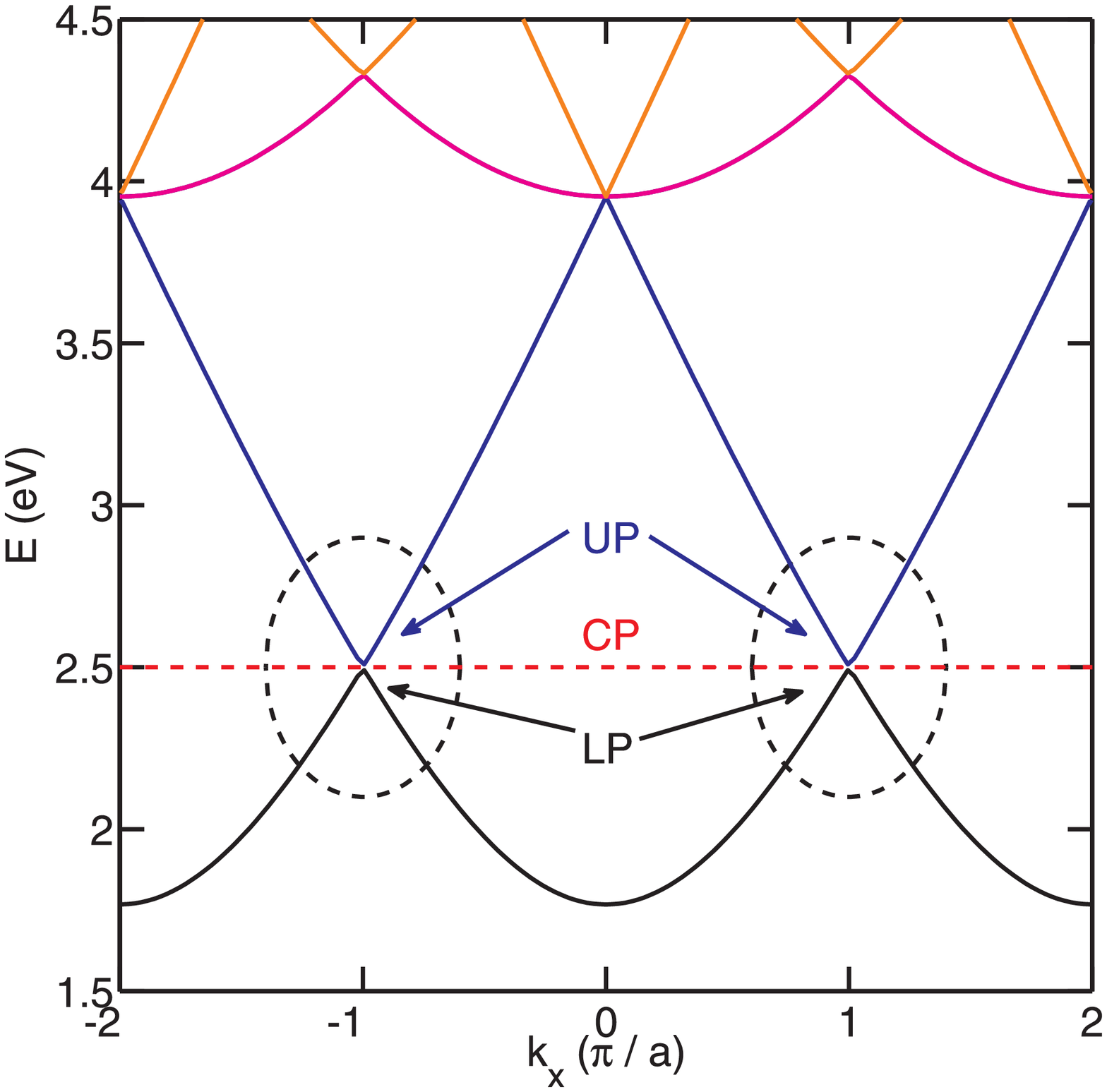}
\caption{(Color online) Scheme of photonic branches (solid lines) in a repeated zone scheme. The exciton energy band is indicated by the flat dashed line. Bragg polaritons are obtained by tuning the exciton energy at the zone boundary where photonic branches cross. Upper, Lower, and Central Polariton modes (indicated by UP, LP, and CP) appear due to the light-matter coupling.}
\label{figbb}
\end{figure}
}

\newcommand{\figc} {
\begin{figure}
\includegraphics[width= 5 cm]{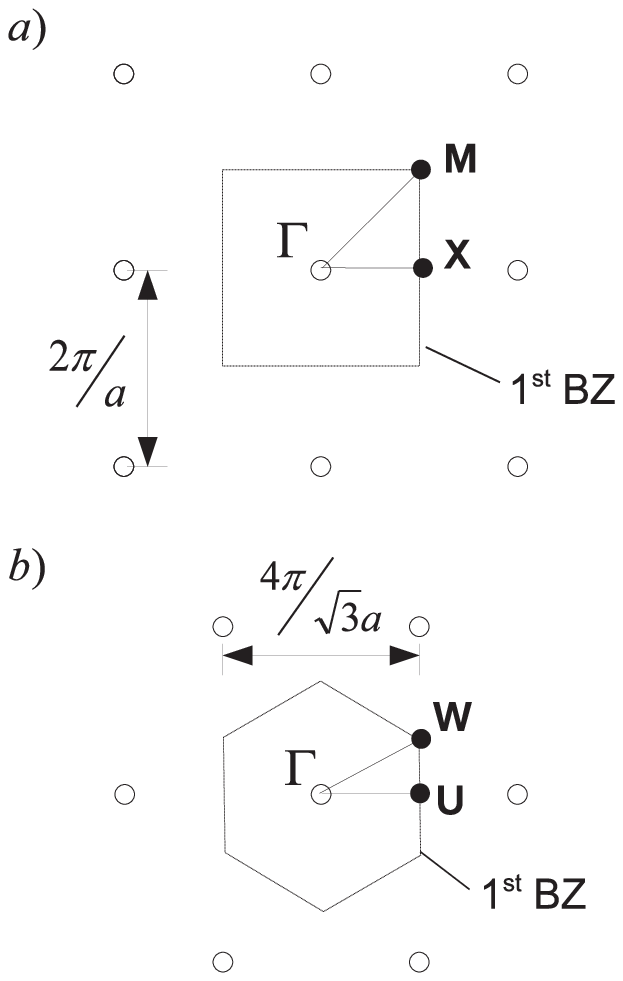}
\caption{Reciprocal lattice of a square lattice (a) and of a
hexagonal lattice with lattice constant $a$ (b). High symmetry points
discussed in the text are indicated.}
\label{figc}
\end{figure}
}

\newcommand{\figd}
{
\begin{figure}
\centering
\includegraphics[width= 5 cm]{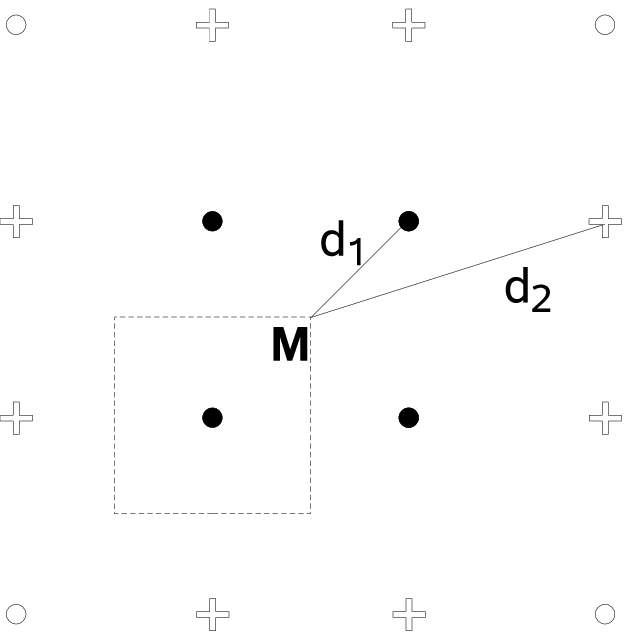}
\caption{Bragg resonances at the M point: M1 labels the energy and
 momentum of Bragg polaritons obtained with the four nearest neighbors
 reciprocal lattice points (black dots). M2 is a higher order Bragg
 resonance involving eight second nearest neighbor reciprocal lattice
 points (crosses).}
\label{figd}
\end{figure}
}

\newcommand{\fige}
{
\begin{figure}
\centering
\includegraphics[width=0.3 \textwidth]{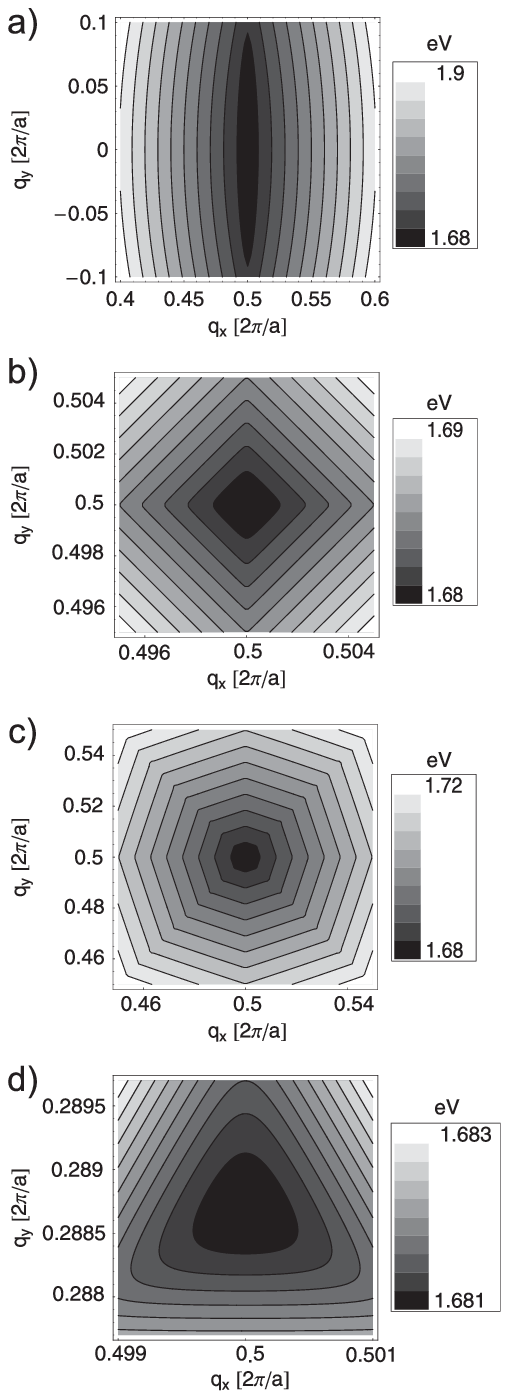}
\caption{Contour plot of the upper polariton dispersion at the
$X$-point (a), $M$-point ($M_1$ in (b) and $M_2$ in (c)), and $W$-point
(d). Note the mass anisotropy in the $X$-point case.}
\label{fige}
\end{figure}
}

\newcommand{\figf} {
\begin{figure*}
\includegraphics[width = 0.9\textwidth, angle=-0]{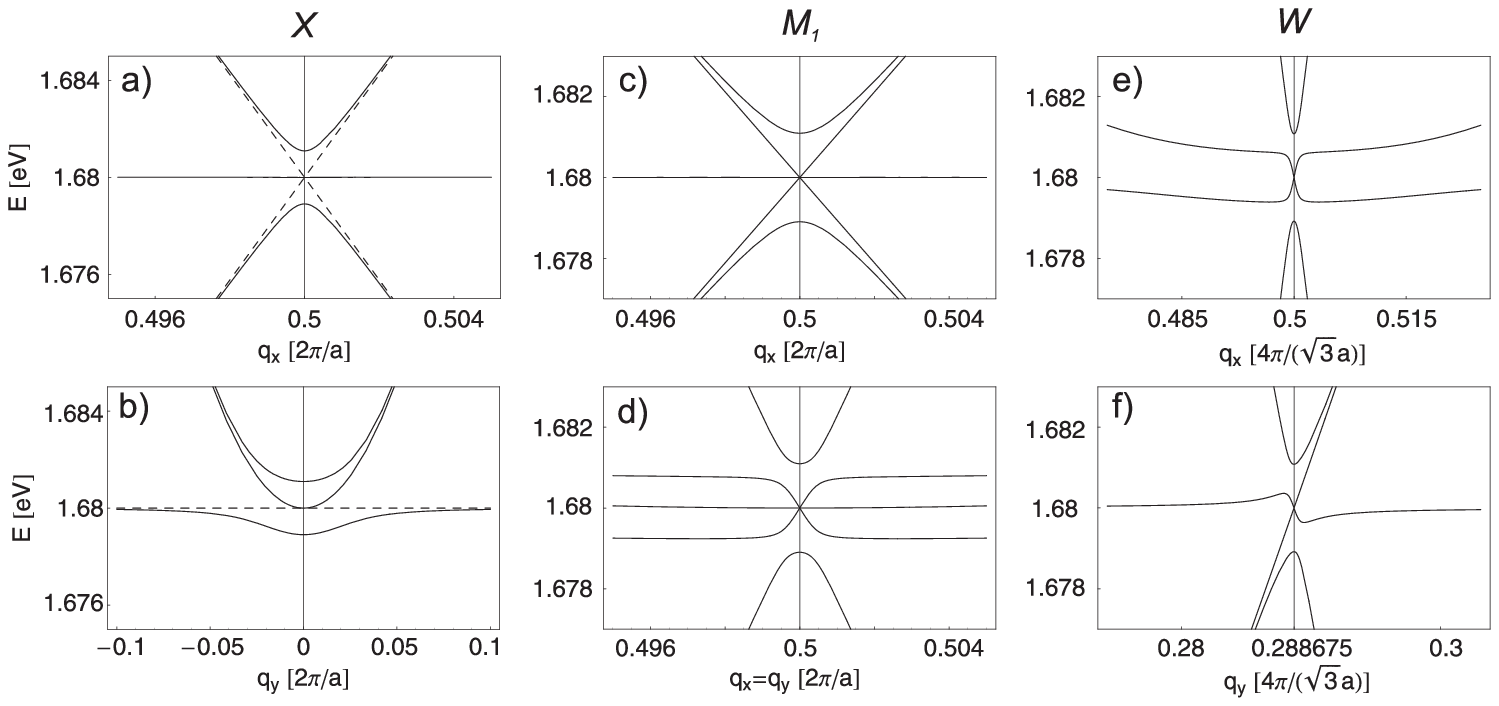}
\caption{The energy dispersion of the Bragg polaritons at selected high symmetry points. $X$-polaritons
along $\overline{\Gamma X}$ (a) and along $\overline{X M}$ (b),
$M_1$-polaritons along $\overline{\Gamma M}$ (c) and $q_x$=$q_y$ axis (d),
and $W$-polaritons along $q_x$-axis (e) and $q_y$-axis (f). The dashed
lines in (a) and (b) represent the uncoupled exciton and photon modes.}
\label{figf}
\end{figure*}
}

\newcommand{\figg} {
\begin{figure}
\centering
\includegraphics[width=0.3 \textwidth]{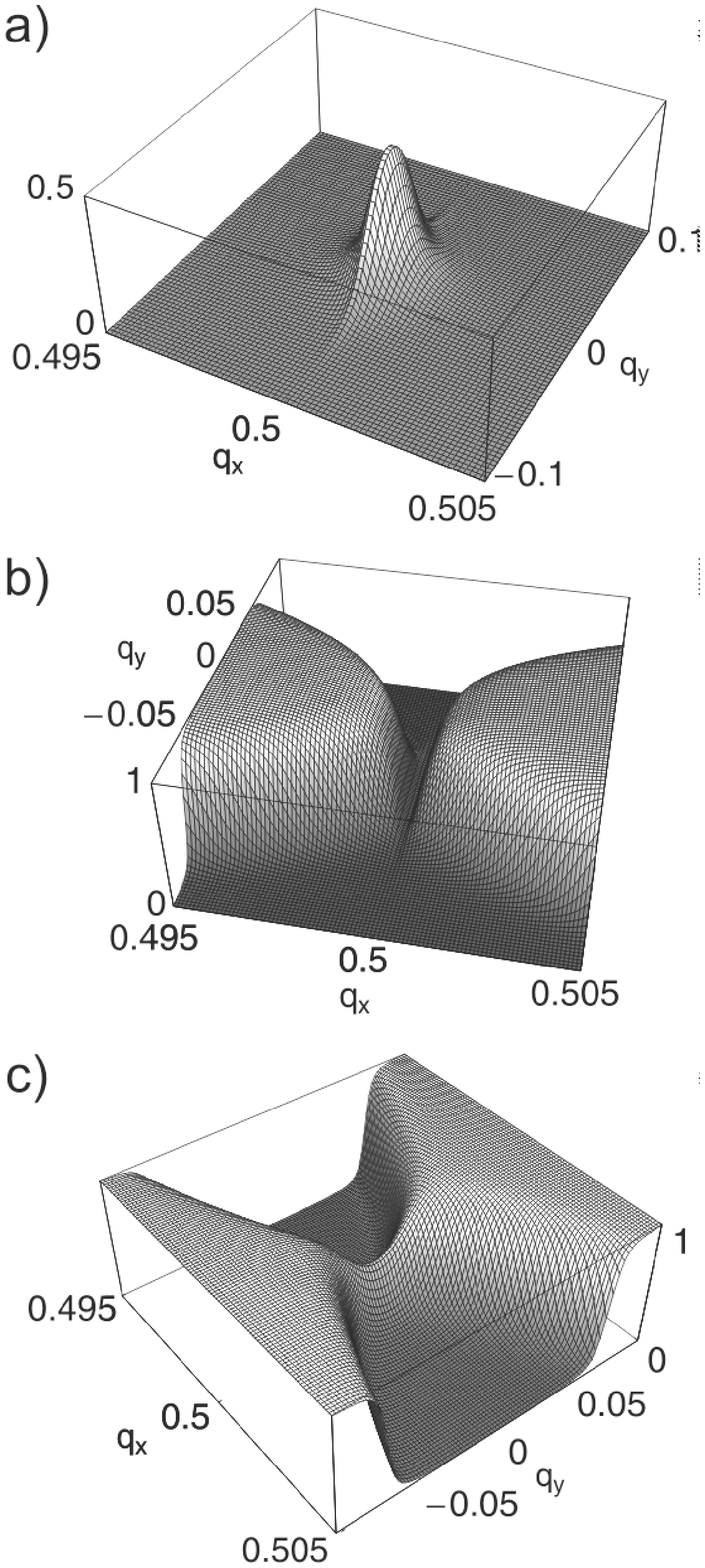}
\caption{The exciton component of the lower (a), central (flat) (b), and
upper (c) Bragg polariton  at the $X$-point.}
\label{fig:figg}
\end{figure}
}

\newcommand{\taba}
{\begin{table}
\caption{Materials parameters used in the numerical calculations}
\label{taba}
\begin{ruledtabular}
\begin{tabular}{ccc}
 & GaAs/AlGaAs & CdSe/ZnSe \\
$\hbar\omega_x$ (eV) & 1.68\footnotemark[1] & 2.50\footnotemark[2] \\
$\hbar\omega^{QW}_x$ (eV) & 1.6 & 2.4 \\
$a^{QD}_B$ (nm) & 11 & 3 \\
$a^{QW}_B$ (nm) & 13 & 6 \\
$g_{QW}$ (meV) & 1.2\footnotemark[3]  & 2.7 \\
$n_R$ & 3.5\footnotemark[4] & 2.5\footnotemark[4] \\
$E_P$ (eV) & 23\footnotemark[5] & 21\footnotemark[6] \\
$\beta$ (nm) & 35 & 35 \\
$L_{DBR}$ ($\mu$m) & 2.0 & 2.0
\end{tabular}
\footnotetext[1]{Ref.~\onlinecite{GSS+1996}.}
\footnotetext[2]{Ref.~\onlinecite{RAG2006}.}
\footnotetext[3]{Comparable with the value found in Ref.~\onlinecite{BJL+2000}.}
\footnotetext[4]{Ref.~\onlinecite{Mad2003}.}
\footnotetext[5]{Ref.~\onlinecite{bastard}.}
\footnotetext[6]{Ref.~\onlinecite{Cardona1963}.}
\end{ruledtabular}
\end{table}
}

\newcommand{\tabb}
{
\begin{table}
\caption{Lower (LP) and upper (UP) polariton effective masses $m_{xx}$
and $m_{yy}$ at $X$- and $M_2$-points for two systems in units of $10^{-5}
m_0$. For comparison, the cavity photon mass $m_{ph}=\hbar n_R
k_z/c$ for the two systems are $2.1~10^{-5} m_0$ and $1.7~10^{-5} m_0$.}
\label{tabb}
\begin{ruledtabular}
\begin{tabular}{ccccc}
& \multicolumn{2}{l}{GaAs/AlGaAs} & \multicolumn{2}{l}{CdSe/ZnSe} \\ &
$X$ & $M_2$ & $X$ & $M_2$ \\ 
\vspace{2mm}
$m_{xx}^{UP}$ & 3.7 $10^{-3}$ & 4.7 $10^{-3}$ & 8.0 $10^{-3}$ & 1.0 $10^{-2}$ \\
\vspace{2mm}
$m_{yy}^{UP}$ & 8.3 & 4.7 $10^{-3}$ & 6.2 & 1.0 $10^{-2}$ \\
\vspace{2mm}
$m_{xx}^{LP}$ & -3.7 $10^{-3}$ & -4.7 $10^{-3}$ & -8.0 $10^{-3}$ & -1.0 $10^{-2}$ \\
\vspace{2mm}
$m_{yy}^{LP}$ & 8.3 & -4.7 $10^{-3}$ & 6.1 & -1.0 $10^{-2}$ \\
\end{tabular}
\end{ruledtabular}
\end{table}
}

\newcommand{\tabc}
{
\begin{table}
\caption{Number of modes $n$ and parameter $\xi$ for the Bragg polaritons discussed in the text.}
\label{tabc}
\begin{ruledtabular}
\begin{tabular}{ccccc}
 & $X$ & $M_1$ & $M_2$ & $W$ \\
$\xi$ & 1 & $\sqrt{2}$ & $\sqrt{10}$ & $4/3$ \\
$n$ & 2 & 4 & 8 & 3 
\end{tabular}
\end{ruledtabular}
\end{table}
}

\newcommand{\tabd}
{
\begin{table}
\caption{Optimal values for the lattice constant $a$ as discussed in
App.~\ref{appb}, and corresponding polariton splitting energies $\hbar
\Omega$ for several high symmetry points for two AlGaAs/GaAs (1) and
CdSe/ZnSe (2) systems. The cavity lenght $L_c$ is $202$ nm and
$182$ nm for the III-V and the II-VI systems, respectively.}
\label{tabd}
\begin{ruledtabular}
\begin{tabular}{ccccc}
 & $X$ & $M_1$ & $M_2$ & $W$ \\
$a_{1}$ (nm) & 121 & 300 & 383 & 161 \\
$a_{2}$ (nm) & 118 & 167 & 373 & 157 \\
$\hbar \Omega_{1}$ (meV) & 2.2 & 2.2 & 1.4 & 2.0 \\
$\hbar \Omega_{2}$ (meV) & 8.8 & 8.8 & 5.6 & 8.0 \\
\end{tabular}
\end{ruledtabular}
\end{table}
}

\begin{document}
\title{Light-mass Bragg cavity polaritons in planar quantum dot lattices}

\author{E. M. Kessler,~\cite{eric} M. Grochol, and C. Piermarocchi}

\affiliation{Department of Physics and Astronomy, Michigan State
University, East Lansing, Michigan 48824 USA}

\date{\today}

\begin{abstract}
The exciton-polariton modes of a quantum dot lattice embedded in a
planar optical cavity are theoretically investigated. Umklapp terms,
in which an exciton interacts with many cavity modes differing by
reciprocal lattice vectors, appear in the Hamiltonian due to the
periodicity of the dot lattice. We focus on Bragg polariton modes
obtained by tuning the exciton and the cavity modes into resonance at
high symmetry points of the Brillouin Zone. Depending on
the microcavity design these polaritons modes at finite in-plane
momentum can be guided and can have long lifetimes. Moreover, their
effective mass can be extremely small, of the order of $10^{-8} m_0$
($m_0$ is the bare electron mass), and they constitute the lightest
exciton-like quasi-particles in solids.
\end{abstract}


\maketitle

\section{Introduction}

The observation of vacuum Rabi oscillations between a quantum dot (QD)
exciton and a cavity photon represents a distinctive example of cavity
quantum electrodynamics realized in a solid state
system.~\cite{reithmaier04,yoshie04,peter05} The strong light-matter
coupling in networks of cavities and dots is very attractive for the
realization of scalable quantum information devices.~\cite{yao05}
Furthermore, fundamental investigations on quantum phase transitions
in extended cavity systems have recently attracted a significant
interest.~\cite{hartmann06, hartmann07, greentree06} The strong
light-matter coupling between excitons and photons is often described
in terms of polariton modes, which identify propagating
electromagnetic modes in a medium with strong dispersive
properties. Lately, there has been a considerable theoretical and
experimental effort on polariton condensation in planar
microcavities.~\cite{deng03,kasprzak06, balili07, keeling07} In the
simplest picture, the critical temperature for polariton condensation
is inversely proportional to the polariton effective mass. Given their
small effective mass (about four orders of magnitude smaller than
excitons), quantum well polaritons suggest the possibility of creating
a condensate at high temperature. Therefore, it is important to investigate
diverse geometrical realizations of polaritonic structures, where the
photon and matter excitations are confined in different ways. These
diverse geometries could lead both to novel quantum information
devices and to novel systems in which coherent matter states can be
obtained at room temperature.

In this paper, we exploit the newly gained opportunity of growing
artificial structures with spatial periodicity comparable to the
wavelength of the confined quasi-particle, the exciton in our case,
and their embedding in a microcavity. The main idea consists in tuning
the exciton energy in such a way that at least two photon modes with
different momentum are resonantly coupled to it due to the lattice symmetry. We assume that the
exciton-photon coupling is stronger than the exciton inhomogeneous
broadening.  There has been many proposals for engineering
exciton-polariton modes using periodic structures: with periodic
quantum well (QW) Bragg structures,\cite{mintsev02,hayes99}
cavity-free three dimensional arrays of QDs,\cite{IFW2000}
point-dipole crystal,\cite{KRP2005} photonic bandgaps with
anti-dots,~\cite{gerace07} and confined QW polaritons in mesa
structures.~\cite{kaitouni06} Here we consider a system with a mismatch between the continuous symmetry of the
two-dimensional photons in the cavity and the discrete symmetry of
a quantum dot lattice. In the proposed structure, the polariton
dispersion is entirely determined by the coupling between the exciton
in the dots and two dimensional photons, which leads to in-plane
effective masses which can be exceptionally small, of order of
$10^{-8} m_0$, for some highly symmetric points in the first Brillouin
zone. The group velocity of 2D photons is large for large in-plane
momenta. This can be seen as the underlying reason behind such a
small polariton mass at the zone edge. The extremely small mass makes this
structure a promising candidate for novel approaches to high
temperature polariton condensation, and for long-range coupling of
spin~\cite{QFP2006}or exciton~\cite{tarel07} qubits.

This paper is organized as follows: In Sec.~\ref{ham} we introduce the
system and write the Hamiltonian for the QD lattice interacting with
the two-dimensional photon modes. The energy dispersion of the
polariton states for some highly symmetric points in the Brillouin
zone of the square and hexagonal lattice is presented in Sec.~III. In
the same section we also discuss the exciton-light mixing, the
effective masses, and the polariton lifetime. Conclusions are found in
Sec.~IV. In the Appendix we provide the explicit form of the
exciton-photon coupling constant, we discuss an optimization procedure
for the system design, and we give the exact effective mass
expressions.

\section{Polariton Hamiltonian}
\label{ham} 
\figa 
We consider the system shown in Fig.~\ref{figa}. A two-dimensional lattice of
QDs or impurities is placed at the center of a planar cavity structure, i.e. a
Fabry-P\'erot resonator. There is currently an intense experimental
effort for controlling the growth of QDs, based on self-assembly and
growth on patterned substrates.~\cite{watanabe05, badolato07}
Controlled impurity implantation or micro-patterned electrostatic
traps on a quantum well~\cite{gartner07,lai07} are further options to
realize the proposed structure. We assume that the array consists of
identical QDs and we use the effective mass, envelope function, and
single sublevel approximations.~\cite{bastard} Furthermore, taking
into account that the confinement potential of a QD can be modelled by
a harmonic potential,~\cite{KLS1990}~ which enables the center-of-mass
and relative motion separation,\cite{Sugawara1995, Sug1997} we write
the ground state heavy hole exciton wave function in the QD centered
at ${\bf R_j}$ in the lattice plane as
\begin{equation}
\label{dotwf}
\Psi_j(\textbf{r}_e,\textbf{r}_h)=\chi(\textbf{R}-\textbf{R}_j)
\Phi(\rho)\phi_e(z_e)\phi_h(z_h)~,
\end{equation}
where $\phi_{e(h)}(z_{e(h)})$ 
is the electron (hole)
$z$-confinement wave function, $\Phi(\rho)=\sqrt{2/\pi a_B^2}
e^{-\rho/a_B}$ is the exciton relative motion wave function with $a_B$
being exciton Bohr radius, and $\chi({\bf R})$ is the exciton
center-of-mass wave function
\begin{equation}
\chi(\textbf{R})=\sqrt{\frac{2}{\pi \beta^2}} \, e^{-R^2/\beta^2},
\end{equation}
where $\beta$ is the effective dot radius. By considering only the ground state we assume that the energy of the first excited state is larger than the exciton-photon coupling, which is of the order of a few meV. 
 
We assume that the photon
modes are confined by ideal planar mirrors in a cavity of length
$L_c$ and we consider only the lowest cavity  mode with the photon dispersion
\begin{equation}
\hbar\omega_\textbf{q}=\frac{\hbar c}{n_R} \sqrt{q^2+k_z^2},
\end{equation}
where $c$ is the speed of light, $n_R$ is refractive index of the
cavity, $k_z$ is fixed to $k_z=\pi/L_c$, and $\textbf{q}$ is the in-plane momentum.  Finally,
the in-plane exciton-photon or polariton Hamiltonian in the second
quantization reads
\begin{eqnarray}
\label{ham1}
H &=& \hbar \omega_x \sum_{j,\mu} \sigma^\dagger_{j,\mu}
\sigma_{j,\mu} + \sum_{\textbf{q},\lambda} \hbar\omega_\textbf{q}
a^\dagger_{\lambda, \textbf{q}} a_{\lambda,\textbf{q}}\nonumber
\\&+&\sum_{j,\mu, \textbf{q},\lambda} \left\{ g^{\lambda*}_{j, \mu, \textbf{q}}
a_{\lambda,\textbf{q}} \sigma^\dagger_{j, \mu}+h.c.\right\},
\end{eqnarray}
where $\hbar \omega_x$ is the exciton energy, $\sigma^\dagger_{j,\mu}$
($\sigma_{j,\mu}$) is the $j^{th}$ QD exciton creation (annihilation)
operator with polarization $\mu$, $a_{\lambda,\textbf{q}}^\dagger$
($a_{\lambda,\textbf{q}}$) is the photon creation (annihilation)
operator for a given polarization $\lambda$. The constant
$g^{\lambda}_{j,\mu,\textbf{q}}$ is the exciton-photon coupling constant
for the $j$-th QD, which has the property
\begin{eqnarray} 
g^{\lambda}_{j,\mu ,\textbf{q}} = e^{i \textbf{q} \cdot \textbf{R}_j}
g^{\lambda}_{0,\mu,\textbf{q}}.
\end{eqnarray}
In the following we are going to focus on TE cavity modes, and we
consider only one exciton polarization. For quantum well polaritons,
TM modes give smaller polariton effects due to the weaker light-matter
coupling,~\cite{savona99} and we expect the same to be true in the
lattice case.  The lattice symmetry can be exploited by introducing
new exciton operators
\begin{equation}
\sigma^\dagger_\textbf{q}=\frac{1}{\sqrt{N}} \sum_j \sigma^\dagger_j
e^{i \textbf{q} \cdot \textbf{R}_j}~.
\end{equation}
After some algebra, the polariton Hamiltonian can be written as
\begin{eqnarray}
\label{ham2}
H &=& \sum_{\textbf{q}\in BZ} \Biggl [ \hbar \omega_x
\sigma^\dagger_\textbf{q} \sigma_\textbf{q} + \sum_{\textbf{Q}}
\hbar\omega_\textbf{q+Q} a^\dagger_{\textbf{q+Q}} a_{\textbf{q+Q}}
\nonumber \\ &+& \sum_{\textbf{Q}} \left \{g_{\textbf{q+Q}}
a_{\textbf{q+Q}} \sigma^\dagger_\textbf{q}+ h.c.\right \} \Biggr ],
\end{eqnarray}
where $\bf{q}$ is now restricted to the first Brillouin Zone ($1^{st}$
BZ) due to the periodicity of the QD lattice and $\bf{Q}$ is a reciprocal lattice vector . The
coupling constant $ g_{\textbf{q}}$ expressed in terms of materials
and structure parameters is given in Appendix~\ref{appa}. 

Moreover, the light-matter interaction conserves the momentum $\bf{q}$ only up to a reciprocal lattice vector.  The
terms in Eq. (\ref{ham2}) involving the exchange of reciprocal lattice vectors are known as exciton Umklapp-processes in the literature.~\cite{bullough70,Knoester1989,KRP2005} This is in contrast to the QW microcavity case where there is a one-to-one correspondence between cavity and exciton modes (the in-plane momentum is conserved exactly).  They can be seen as interband scattering in which a valence electron $e_v$ with momentum
$\textbf{q}_1$ scatters with a photon with momentum $\textbf{q}_2$ and creates
a conduction electron $e_c$. If the final momentum
$\textbf{q}_1+\textbf{q}_2$ lies outside the $1^{st}$ BZ, then the
momentum of the conduction electron is flipped back to the first BZ by
adding a reciprocal lattice vector.
Since the photon momentum $\textbf{q}_2$ is not restricted to the
$1^{st}$ BZ, the Umklapp momentum $\textbf{Q}$ for excitons can have
an arbitrary large magnitude. This is a difference with respect to
electron-phonon Umklapp processes,~\cite{AshcroftMermin} where only the
first neighbor reciprocal lattice points are involved.
 
However, we remark that in Eq.~(\ref{ham2}) the quantum
dot-photon coupling has a structure form factor due to the finite size
of the quantum dots, i. e.
\begin{eqnarray}
g^{\lambda}_{\textbf{q}} \propto e^{-\beta^2 q^2/4}~,
\end{eqnarray} 
so that we find a natural cut-off for the reciprocal lattice vectors. Furthermore, in the photonic case, excitonic corrections due to the Coulomb interaction between the conduction electron and the hole left in the valence band have to be included.

\figbb
\figc 
For an exciton state with momentum $\textbf{q}$ away from the zone
boundary, many off-resonant Umklapp terms give corrections which do
not entail qualitative novel properties for the polariton
quasi-particles.  However, the situation can be very different if the
structure is built in such a way that the exciton is resonant with the
cavity at a $\textbf{q}_0$ which is at or near the BZ boundary.  In
that case we can choose, for instance, $\textbf{Q}_i$ and
$\textbf{Q}_j$ so that the Bragg condition
\begin{eqnarray}
\label{Bragg}
\omega_x \sim \omega_{\textbf{q}_0-\textbf{Q}_{i}} \sim
\omega_{\textbf{q}_0-\textbf{Q}_{j}}
\end{eqnarray}
is satisfied. In general, the number $n$ of reciprocal lattice points
satisfying this condition depends on the lattice symmetry and on the
energy of the exciton. The mode configuration leading to the formation
of Bragg cavity polaritons is illustrated in Fig.~\ref{figbb}, where
the exciton energy (red dashed horizontal line) is tuned in such a way
that the polariton mixing occurs at the zone boundary. Due to the
reduced symmetry of the dot lattice the $k_z=\pi/L_c$ cavity mode is
folded giving rise to many photonic branches, each characterized by a different reciprocal lattice vector ${\bf Q}$. In order to visualize this folding, the figure shows two repeated BZs. 
Bragg polariton modes will appear at $\textbf{q}_0$ where two or more photonic branches are crossing and are nearly degenerate with the exciton energy. These polaritons will have
different properties than polaritons usually investigated at $\textbf{q}_0\sim 0$. Fig.~\ref{figc} shows the high symmetry points at the zone boundary for the square and hexagonal lattice for which we will discuss Bragg polaritons in the following.

Due to the cut-off introduced by the finite quantum dot size, we can
write the single polariton Hamiltonian at fixed $\textbf{q}$ in the form
\begin{equation}
\label{tmatrix}
\begin{pmatrix}
\hbar \omega_x &g_{\textbf{q}} & g_{\textbf{q}+\textbf{Q}_1} &
g_{\textbf{q}+\textbf{Q}_2} &\cdots& g_{\textbf{q}+\textbf{Q}_{max}}\\
g_{\textbf{q}} &\hbar \omega_{\textbf{q}}& 0 & 0 & \cdots & 0\\
g_{\textbf{q}+\textbf{Q}_1} & 0 &\hbar
\omega_{\textbf{q}+\textbf{Q}_1}& 0 & \cdots & 0\\
g_{\textbf{q}+\textbf{Q}_2} & 0 & 0 & \hbar
\omega_{\textbf{q}+\textbf{Q}_2} &\cdots& 0\\ \vdots & \vdots &
\vdots & \vdots & \ddots& \vdots\\ g_{\textbf{q}+\textbf{Q}_{max}} & 0 &
0 & 0 & \cdots & \hbar \omega_{\textbf{q}+\textbf{Q}_{max}}
\end{pmatrix}~.
\end{equation}
Furthermore, the finite size of the dots limits the largest reciprocal lattice vector to $|\textbf{Q}_{max}| \sim 2 \pi/\beta$. If we keep only the $n$ resonant terms at the zone boundary satisfying
exactly the Bragg condition, the matrix Eq.~(\ref{tmatrix}) can be
diagonalized analytically and we find the two eigenvalues
\begin{equation}
\lambda_{1,2}=\hbar \omega_x \pm \sqrt{n} g_{\textbf{q}_0}~,
\end{equation}
corresponding to the strongly mixed polariton states, as well as the
($n-1$)-fold degenerate eigenvalue
\begin{equation}
\lambda_3=\hbar \omega_x~.
\end{equation}
The normal mode splitting (Rabi energy), i. e. the difference between the lowest
(lower polariton) and the highest (upper polariton) energy at
resonance, is then 
\begin{equation}
\label{rabi}
\hbar \Omega = 2\sqrt{n} g_{\textbf{q}_0},
\end{equation}
and is proportional to the square root of the number of reciprocal
lattice vectors involved in the Bragg condition.  

In the square lattice, we focus on Bragg polaritons obtained at the
$X$ and $M$ high symmetry points defined in Fig. \ref{figc} (a). The
number of photonic branches that can be brought into resonance is
determined by the number of points in the reciprocal lattice
equidistant to $X$ or $M$, according to Eq.~(\ref{Bragg}).  There are
two nearest equidistant points to $X$ ($n=2$) and four nearest
equidistant points to $M$ ($n=4$) (see dots in
Fig.~\ref{figd}). However, if we do not concentrate on the lowest
photonic branches and the nearest-neighbors in the reciprocal lattice, but
e.g. on the second nearest neighbors, then a larger number of reciprocal
lattice vectors can be involved. For instance, there are eight
second-nearest neighbor equidistant points to $M$ ($n=8$, see crosses
in Fig.~\ref{figd}). In the following we will label the $M$ polariton
states as $M_1$ if $n=4$ and $M_2$ if $n=8$. For the hexagonal lattice
(see Fig.~\ref{figc}(b)) we restrict our discussion to the $W$-point,
which has $n=3$. The $U$-point is similar to the $X$-point of the
square lattice and will not be discussed. Since it is possible to express
explicitly the coupling constant for a given high symmetry point of
length $q_0=\xi \pi/a$ (see Eq. (\ref{galpha}) in
Appendix~\ref{appb}). Then the normal mode splitting can be written as
\begin{equation}
\hbar \Omega = 2 \frac{\sqrt{n}}{\xi} g_X,
\end{equation}
where $g_X$ is the coupling constant at the $X$-point. The values of
$n$ and $\xi$ can then be used as scaling parameters to calculate the strength of polaritonic effects at different symmetry points, and  are summarized in
Tab. \ref{tabc}.
\figd 
\tabc

\subsection{Effective mass at X-point}

One of the most interesting features of Bragg polaritons consists in
their extremely small effective mass, which is a consequence of the
fact that two-dimensional photons at large momentum enter in the
polariton formation. In this section, we investigate this property
analytically for a Bragg polariton at the $X$-point. The photon energy in
the vicinity of the $X$-point can be expanded to the second order in the
in-plane momentum as
\begin{eqnarray}
E_X(q_x,q_y) &=& \frac{\hbar c}{n_R} \sqrt{k_z^2 +
(\pi/a-q_x)^2+q_y^2} \nonumber \\ &\sim&  E_X^0 + \alpha (-2
\pi q_x/a + \nu^2 q_x^2 + q_y^2) \nonumber
\end{eqnarray}
where $E_X^0$ is the $X$-point photon energy, $\alpha = E_X^0/2 K^2$ and
$\nu = k_z/K$ with $K = \sqrt{k_z^2 + (\pi/a)^2}$.
Taking only into account the exciton modes and the two photonic
branches nearly resonant in the vicinity of the $X$-point, the full Hamiltonian
in Eq.~(\ref{tmatrix}) matrix can be simplified as
\begin{equation}
\begin{pmatrix}
0 & g_X  & g_X \\
g_X & \alpha (-2 \pi q_x/a + \nu^2 q_x^2+ q_y^2) & 0 \\
g_X & 0 & \alpha (2 \pi q_x/a + \nu^2 q_x^2+q_y^2)
\end{pmatrix}\nonumber
\end{equation}
where the zero of the energy is set at $E_X^0$.  The eigenvalues of this
matrix can be found analytically by solving a cubic equation. From
the expansion of the eigenvalues to the second order in $q_x$ and
$q_y$ we obtain
\begin{eqnarray}
\frac{E_{UP(LP)}}{E_X^0} &=& 1 \pm \sqrt{2} \tilde{g}_X + \frac{1}{4
K^2}\left( q_y^2 + \nu^2 q_x^2 \right ) \nonumber
\\ & \pm& \frac{\sqrt{2}}{4 K^2 \tilde{g}_X } \left ( \frac{\pi}{a K}
\right)^2 q_x^2,\\ \frac{E_{CP}}{E_X^0} &=& 1 +
\frac{1}{2 K^2} \left(q_y^2 +\nu^2 q_x^2\right)~,
\end{eqnarray}
where the $UP$, $CP$, and $LP$ index  identify the upper, central, and
lower polariton branches, respectively. The upper
(lower) sign refers to UP (LP), and
\begin{eqnarray}
\tilde{g}_X = \frac{g_X}{E_X^0}, 
\end{eqnarray}
is the renormalized coupling constant. From the
definition of the effective mass tensor it
follows for the upper and lower polariton branches
\begin{eqnarray}
\label{efmas2}
\frac{1}{m_{xx}} &=& \frac{E_X^0}{2 \hbar^2 K^2} \left [ \nu^2 \pm \frac{\sqrt{2}}{\tilde{g}_X } \left
( \frac{\pi}{a K} \right )^2 \right] \nonumber \\
\frac{1}{m_{yy}} &=&
\frac{E_X^0}{2 \hbar^2 K^2}.
\end{eqnarray}
In these expressions, we have neglected the weak dependence of the
coupling constant $g_X$ on the momentum $\bf{q}$. This dependence gives small
corrections to the effective mass and is discussed in
Appendix~\ref{appc}. Since typically $\tilde{g}_X \sim 10^{-3}$, and
$L_c \sim a$, the effective mass ratio reads
\begin{eqnarray}
\label{massratio}
\frac{m_{xx}}{m_{yy}} = \frac{\tilde{g}_X}{\sqrt{2}} \left (
\frac{K a}{\pi} \right )^2 \sim 10^{-3}~.
\end{eqnarray}
Therefore, the polariton mass is very anisotropic at the $X$-point.  We
can also compare the value of these effective masses with the QW
polariton mass, which is of the same order of magnitude of the cavity
photon mass at $q \sim 0$
\begin{eqnarray}
\label{phmass}
m_{ph} = \frac{\hbar n_R}{c} k_z,
\end{eqnarray}
and typically $\sim 10^{-5} m_0$.  Then, we remark that for the the
mass along the $\Gamma X$ direction we have $m_{xx}/m_{ph} \sim
10^{-3}$ and $m_{yy}/m_{ph} \sim 1$. This implies that a value of the order of $10^{-8} m_0$ is
expected. Moreover, we will see in the next section that isotropic masses of such a order can be obtained at special high symmetry points.

\section{Results and discussion}
\tabd 
\fige 
\figf 
In this section we present the numerical results for three high
symmetry points of the square and hexagonal lattice. The numerical
calculations include all resonant and off resonant terms.  We have
taken parameters from AlGaAs/GaAs and CdSe/ZnSe systems, which are
typical examples of III-V and II-VI quantum dot structures. In
general, II-VI semiconductors microcavities have a  stronger
light-matter coupling, e.g. for CdTe microcavity the light-matter
coupling is $g \sim 20$ meV,\cite{AHD+1998} while for III-V systems the
coupling constant is $g \sim 2$ meV.\cite{SAS+1995} The cavity length,
dot size, and lattice constant can be optimized in order to maximize the
light-matter coupling for Bragg polaritons at a given high symmetry
point.  The optimization procedure is discussed in the
Appendix~\ref{appb} and the parameters are given in Tab.~\ref{tabd}.

Fig.~\ref{fige} ($a$-$d$) shows contour plots of the energy
dispersion of the upper polariton branch for Bragg polaritons at the
$X$-point ($a$), $M$-point ($b$ for $M_1$-polariton and $c$ for $M_2$-polariton) of the square lattice, and $W$-point ($d$) of the hexagonal
lattice.  The energy dispersion near the resonance reflects the
symmetry of these points. We can also see that the polariton
effective mass is highly anisotropic for the $X$-polariton, as
discussed in the previous section, while the effective mass for $M_2$
is almost isotropic. In Fig.~\ref{figf} we show the energy dispersion of the
Bragg polariton modes for $X$-polaritons, $M_1$-polaritons, and $W$-polaritons. The
different behavior of the UP, CP, and LP modes at different points
shows how richer Bragg polaritons are compared to the QW polaritons. Note
that the total number of branches crossing or anti-crossing at the
different symmetry points is always given by $n+1$. 

At the $X$-point, we have $n=2$ and consequently there are three polariton modes. In the $q_y$-direction, the upper and lower $X$-polaritons are similar to the QW polariton (see Fig. \ref{figf}b). However, there is one additional
central mode degenerate with the photon mode along the zone boundary in the $q_y$-direction. On the contrary, the $q_x$ dependence along $\overline{\Gamma X}$ shows the existence of (i) a lower branch with a
negative effective mass in one direction, (ii) an upper branch with
positive effective masses but high anisotropy, and (iii) a flat branch
with a large mass (consequence of the infinite exciton mass considered
in our model) with energy exactly between the lower and upper branch. 

Let us now look at the $M_1$-polaritons, which have five branches. In the $q_x=q_y$ direction we can
see nearly flat photonic branches originating from photon modes with $\textbf{Q}=(2\pi/a,0)$ and $\textbf{Q}=(0,2 \pi/a)$, while the polariton mixing occurs mainly with photon modes having $\textbf{Q}=(0,0)$ and
$\textbf{Q}=(2\pi/a,2 \pi /a)$. A similar plot for $M_2$ (not shown)
gives nine polariton branches. Notice that there is not a saddle point at the $M$-point. Polariton mode with isotropic negative mass (lower branch) can give rise to an effective negative index of
refraction.~\cite{agranovich06} In the $W$-point case there are four branches and the properties are similar to the $M$-point of the square lattice.

In the case of QW polaritons, the exitonic component is equally shared between lower and upper polaritons at the anticrossing.  This is also true for the excitonic component in the $X$-point as shown in Fig. \ref{fig:figg} ($a$ and $c$). At the anti-crossing, the UP and LP branches consist of {\it half exciton} and consequently, the excitonic component is
completely absent in the CP branch. Nevertheless, it increases rapidly away from the anticrossing region. This exciton-photon
character swap is a characteristic of the Bragg polaritons since in all the cases investigated we have found analytically and numerically that the excitonic component is always equally shared between the lower and upper polaritons at the anticrossing, the rest being purely photonic.

\figg

\tabb

The effective masses calculated numerically for the two material systems are given in Table~\ref{tabb}. As expected a strong anisotropy at the $X$-point and almost isotropic mass at $M_2$-point are found. The numerical values for the $X$-point are the same as calculated using the analytical formula Eq. (\ref{efmas2}), thus validating our numerical approach. Furthermore, the ratio in Eq. (\ref{massratio}) for both AlGaAs/GaAs and CdSe/ZnSe systems, for which $\tilde{g}_X \sim 10^{-3}$ and $\tilde{g}_X \sim 10^{-2}$, respectively (see Appendix ~\ref{appb} for materials parameters) is also obtained. The value of the effective mass in some cases is of the order of $10^{-8}m_0$ and is extremely small for a matter like quasi-particle. As discussed in the Introduction, the small effective mass could suggest new strategies for the realization of polariton condensation in microcavities. However, the dynamics of Bragg polaritons, including the phonon bottleneck effects,\cite{TPS+1997} and polariton-polariton scattering is expected to be different than for quantum well polaritons, and needs further investigations. The small polariton effective mass can play an important role also in quantum information implementations. It has been recently shown that QW polaritons can mediate a spin-spin coupling between two charged dots or impurities over a distance of the order of hundreds of nanometers due to the light mass of polariton.\cite{QFP2006} Since Bragg polariton states have effective masses that are two to three orders of magnitude smaller, the range of the spin coupling could be further extended.

So far, the lifetime of Bragg polaritons has not been discussed. There is an essential difference with respect to the QW polariton case, where the light matter mixing occurs close to $q=0$.  The quantum well polariton lifetime depends on the photon lifetime in the cavity, which is determined by the reflectance of the mirrors in the normal direction and is of the order of picoseconds. However, at finite $\textbf{q}$, depending on the external index of refraction and on the propagation angle $\theta_p=\arccos(\frac{k_z}{K})$, a cavity photon can either be transmitted through the mirrors, or it can remain confined due to total internal reflection. This means that structures can be engineered where the polariton mixing occurs at large $\textbf{q}$ with {\it guided} modes, which is never possible in the quantum well case. Since guided photons have in principle a much longer lifetime, the polariton lifetime will also be much longer, likely in the order of micro or milliseconds, depending on the properties of the lateral surfaces of the cavity structures. However, the details of the cavity modes at large in-plane momentum, in particular leaky modes~\cite{TPS+1996} due to the distributed Bragg mirrors, need to be carefully taken into account, since they will affect both the dynamics and the radiative properties of these polaritons.

\section{Conclusions and outlook}
\label{conclusions}

We have investigated polariton modes in a planar quantum dot lattice embedded in a microcavity. Assuming small exciton and photon homogeneous broadening, we have considered systems in the strong coupling regime. We have studied Bragg polariton modes formed at high symmetry points of the Brillouin Zone. Truly polaritonic quasi-particles with a fifty-fifty ligh-matter character appear at these points in upper and lower branches. Analytical and numerical calculations of the polariton effective masses give extremely small values, of the order of $10^{-8}m_0$, and in some cases strong anisotropy.  The small mass is due to the mixing of excitons with nearly linear two-dimensional photon modes at large momenta. There is a full transfer of the excitonic wavefunction component from the flat exciton dispersion to photonic branches in the strong coupling region. We have also discussed the possibility of polariton coupling with guided modes, which could lead to long polariton lifetime. 

Future extensions of this work could include: (i) an explicit theoretical treatment of the homogeneous broadening with a discussion of the strong to weak coupling transition, (ii) realistic structure modelling of the distributed Bragg mirrors  with leaky modes, which could be close to the Bragg polariton resonances, and (iii) sensitivity to inhomogeneous broadening effects. The polariton-phonons, polariton-polaritons, and polariton-spin dynamics for these Bragg quasi-particles also needs further investigation, since it is expected to be very different than in the  quantum well case.

\begin{acknowledgments}
This research was supported by the National Science Foundation, Grant
No. DMR-0608501.
\end{acknowledgments}

\appendix

\section{Coupling constant}
\label{appa}

The lattice exciton-photon coupling in the cavity for the TE mode
reads\cite{savona99}
\begin{equation}
\label{qdcoupl}
g^{TE}_{\textbf{q}} = i e \omega_x \sqrt{\frac{\hbar}{\epsilon_0
n_R^2 \omega_{\textbf{q}}V}} u_{cv}
\sqrt{N} \Phi_{1s}(0) \tilde{\chi}(\textbf{q}) F(L_c),
\end{equation}
where $V = S L_c $ is the volume of the cavity, $S$ being the
surface of the cavity, $u_{cv}$ is the dipole matrix element between
conduction and valence band, $N$ is the number of dots in the cavity,
$\tilde{\chi}(\textbf{q})$ is the center-of-mass form factor
\begin{eqnarray}
\tilde{\chi}(\textbf{q}) &=& \sqrt{2 \pi} \beta e^{-\textbf{q}^2
\beta^2/4}, \nonumber
\end{eqnarray}
and $F(L_c)$ is the $z$-form factor 
\begin{equation}
F(L_c) = \int dz \phi_e(z)\phi_h(z)\cos \left(\frac{\pi
z}{L_c}\right).\nonumber
\end{equation}
We can assume that the QW that contains the quantum dots is much
narrower than the cavity length $L_c$, which implies $F(L_c)\sim 1$. The
coupling constant for TM modes
\begin{equation}
g^{TM}_{\textbf{q}} = \frac{-ik_z}{\sqrt{q^2+k_z^2}} g^{TE}_{\textbf{q}}~,
\end{equation}
is always smaller than for TE modes at finite in-plane momentum.  We
can relate the QD coupling constant in Eq.~(\ref{qdcoupl}) to the
equivalent quantity in the QW case as
\begin{equation}
g^{TE}_{\textbf{q}} = \frac{\hbar \omega_x}{\hbar \omega_x^{QW}} \, \sqrt{n_D}
\, \tilde{\chi}_0(\textbf{q}) \frac{a_B^{QW}}{a^{QD}_B} g^{TE}_{QW},
\end{equation}
where $\hbar \omega_x^{QW}$ is a QW exciton energy, $n_D=N/S_c$ is
the QD density, $a_B^{QW}$ ($a_B^{QD}$) is a Bohr radius of the QW (QD) exciton. We note that recent numerical calculations show that for confined excitons the effective Bohr radius  defined by the ansatz of Eq.~(\ref{dotwf}) is typically smaller that the quantum well value.~\cite{grochol05,GGZ2005} The quantity $n_{QD} 2 \pi \beta^2$ can be understood as a filling or packing factor. For a fixed dot size and exciton energy the hexagonal lattice has the largest packing factor and the
largest coupling at $\textbf{q}=0$.  The value of the interband matrix element can be calculated conviniently from the Kane
energy parameter using the relation\cite{RosencherVinter2002}
\begin{equation}
u_{cv}^2 = \frac{\hbar^2 e^2}{\left ( \hbar\omega^{QW}_x \right)^2} \frac{E_P}{2 m_0}.
\end{equation}
In order have a strong light matter coupling, a system with a large
dipole matrix element and with small QD Bohr radius is desired. From this
point of view II-VI semiconductors are better candidates than III-V. Our
scheme could be also implemented with impurity-bound excitons, which
usually have very strong oscillator strength, large spatial extension,
and very small inhomogeneous broadening.~\cite{SAL+1996} The materials
parameters used in the calculations are given in Tab.~\ref{taba}. 

\section{Parameter optimization}
\label{appb}
\taba 

Let us now fix the QD size and ground state exciton energy. The lattice
constant $a$ and cavity length $L_c$ are then adjustable parameters
which can be optimized to increase the polariton mixing. The coupling
constant at a high symmetry point with $q_0 = \xi
\frac{\pi}{a}$ can be written as
\begin{equation}
\label{coupling2}
g_{q_0}(a,L_c,\xi)=
\frac{1}{\sqrt{L_c+L_{DBR}}} \frac{\beta}{a} e^{-\frac{\beta^2 \xi^2 \pi^2}{4 a^2 } } g_c,
\end{equation}
where $g_c$ is a constant which does not depend on $\beta$, $a$, and
$L_c$. Note that the dependence on the lattice constant $a$ is non-monotonic due to the competition between the quantum dot density and the structure factor. Moreover, the exciton-photon Bragg resonance condition gives the constraint
\begin{equation}
\label{constrain}
\hbar \omega_x = \frac{\hbar c}{n_R} \sqrt{\left( \frac{\xi \pi}{a} \right)^2 +
\left( \frac{\pi}{L_c} \right)^2}.
\end{equation}
In order to take into account that distributed Bragg reflectors are typically used as mirrors, we have added an effective $L_{DBR}$ to the nominal cavity spacer length $L_c$ in the coupling constant.~\cite{SAS+1995} By expressing $L_c$ as a function of $a$ from Eq.~(\ref{constrain}), the coupling constant becomes a function of only one variable $g_{q_0}(a)$. The optimal lattice spacing can then be found by analytical or numerical maximization. Defining $a'=a/\xi$, we obtain from Eq.~(\ref{coupling2})
\begin{equation}
\label{galpha}
g_{q_0} (a,L_c,\xi)= \xi^{-1} g_{q_0}(a',L_c,1).
\end{equation}
Thus one needs to optimize $a$ only for one high symmetry point of each material combination and then rescale it by $\xi$ for other points.  

\section{Effective Mass Corrections}
\label{appc}

Here we include the corrections to the effective mass due to the dependence of the coupling constant on $\textbf{q}$. This provides the exact second order expansion in $q_x$ and $q_y$ of Bragg polaritons and therefore the exact expression for their effective mass tensor. Only $g^{TE}_{\textbf{q}} \equiv g_X$ close to the $X$-point is considered here. The coupling constant written as a function of small ${\bf q}$ around the $X$-point reads
\begin{equation}
g_X(q_x, q_y) = g^0_X \frac{e^{-(2  \pi q_x/a + q_x^2 + q_y^2)
\beta^2/4}}{\sqrt[4]{1 + (2 \pi q_x/a + q_x^2 + q_y^2)/K^2}}~.\nonumber
\end{equation}
Using $(1+x)^{-\frac{1}{4}} = 1 - \frac{1}{4} x + \frac{5}{32} x^2 + O(x^3)$ and the expansion of the exponential function we obtain
\begin{equation}
\label{gexpan}
g_X(q_x, q_y) = g^0_X + g^{1,x}_X \frac{q_x}{K} + g^{2,x}_X \left (
\frac{q_x}{K} \right )^2 + g^{2,y}_X \left ( \frac{q_y}{K} \right )^2~,
\nonumber
\end{equation}
where
\begin{eqnarray}
 g^{1,x}_X &=& - g^0_X \frac{\pi}{2 a} \left ( \frac{1}{K} + K \beta^2
\right ), \nonumber \\ g^{2,x}_X &=& g^0_X \left [ -\frac{1}{4} \left
(1 + K^2 \beta^2 \right ) + \frac{5}{8} \left(\frac{\pi}{K a}\right)^2
\right ], \nonumber \\ g^{2,y}_X &=& g^0_X \left [ -\frac{1}{4} \left
(1 + K^2 \beta^2 \right ) \right ]~.\nonumber
\end{eqnarray}
The exact expansion for the polaritonic modes is therefore
\begin{widetext}
\begin{eqnarray}
\frac{E_{UP(LP)}}{E^0_X} &=& 1 \pm \sqrt{2} \tilde{g}^0_X + \left [
\frac{1}{4} \pm \sqrt{2} \tilde{g}_X^{2,y} \right ] \left (
\frac{q_y}{K} \right )^2 + \left [\left ( \frac{\nu}{2} \right )^2 \pm
\frac{1}{2 \sqrt{2} \tilde{g}_X^0 } \left ( \frac{\pi}{K a} \right )^2
\pm \sqrt{2} \tilde{g}_X^{2,x} \mp \frac{1}{\sqrt{2}}
\frac{(\tilde{g}_X^{1,x})^2}{\tilde{g}_X^0} - \frac{g^{1,x}_X}{g^0_X} \frac{2
\pi}{K a} \right ] \left ( \frac{q_x}{K} \right )^2, \nonumber \\
\frac{E_{CP}}{E_X^0} &=& 1 + \frac{1}{2} \left [ \nu^2 +
\frac{g^{1,x}_X}{g^0_X} \frac{4 \pi }{K a} \right ] \left ( \frac{q_x}{K}
\right )^2 + \frac{1}{2} \left ( \frac{q_y}{K} \right )^2~,
\end{eqnarray}
The effective masses for the UP(LP) can be written as
\begin{eqnarray}
\label{efmas}
\frac{1}{m_{xx}} &=& \frac{E_X^0}{\hbar^2 K^2} \left [ \frac{\nu^2}{2}
\pm \frac{1}{\sqrt{2} \tilde{g}_X^0 } \left ( \frac{\pi }{K a} \right
)^2 \pm \sqrt{8} \tilde{g}_X^{2,x} \mp \sqrt{2}
\frac{(\tilde{g}_X^{1,x})^2}{\tilde{g}_X^0} \pm \frac{g^{1,x}_X}{g^0_X}
\frac{4 \pi }{K a} \right ] \nonumber \\ \frac{1}{m_{yy}} &=&
\frac{E_X^0}{\hbar^2 K^2} \left ( \frac{1}{2} \pm \sqrt{8}
\tilde{g}_X^{2,y} \right )~,
\end{eqnarray}
\end{widetext}
while for the CP we have
\begin{eqnarray}
\frac{1}{m_{xx}}&=&\frac{E^0_X}{2 \hbar^2 K^2}\left[\nu^2+\frac{g^{1,x}_X}{g^0_X}\frac{4 \pi}{K a}\right] \nonumber \\
\frac{1}{m_{yy}}&=&\frac{E^0_X}{2 \hbar^2 K^2}~.
\end{eqnarray}
The upper and lower polaritons small mass in the $x$-direction is mainly due to the $O(1/\tilde{g_0})$ term and the $\textbf{q}$ dependence of the coupling constant provides only small corrections. The numerical calculations of the effective mass given in Table~\ref{tabb} include the explicit dependence of the coupling constant on the wavevector.

\bibliographystyle{apsrev}

\end{document}